# M-AODV: AODV variant to Improve Quality of Service in MANETs

Maamar Sedrati [1], Azeddine Bilami[2] and Mohamed Benmohamed[3]

[1] UHL Batna, Department of computer science
Batna, Algeria

[2] UHL Batna, Department of computer science
Batna, Algeria

[3] UMC Constantine, Department of computer science
Constantine, Algeria

**Abstract**
Nowadays, multimedia and real-time applications consume much network resources and so, need high flow rates and very small transfer delay. The current ad hoc networks (MANETs), in their original state, are not able to satisfy the requirements of quality of service (QoS). Researches for improving QoS in these networks are main topics and a subject of intensive researches.
In Adhoc networks, the routing phase plays an important role for improving QoS. Numerous routing protocols (proactive, reactive and hybrid) were proposed. AODV (Adhoc On demand Distance Vector) is probably the more treated in literature
In this article, we propose a new variant based on the AODV which gives better results than the original AODV protocol with respect of a set of QoS parameters and under different constraints, taking into account the limited resources of mobile environments (bandwidth, energy, etc…).
The proposed variant (M-AODV) suggests that the discovering operation for paths reconstruction should be done from the source. It also defines a new mechanism for determining multiple disjoint (separated) routes.
To validate our solution, simulations were made under Network Simulator (NS2). We measure traffic control and packet loss rate under diverse constraints (mobility, energy and scale).

## 1. Introduction

The new multimedia applications (videoconferencing, video telephony, web games, etc.) and real-time require respectively high throughput and reduced delays that are among fundamental quality of service (QoS) parameters.
Providing quality of service for networks, particularly for Adhoc networks (MANETs), have and still being the subject of intensive research in order to propose better solutions for these new requirements, not only in particular level but at different levels of network architecture i.e. by various network layers (physical, network, etc.) [1] [2].
The routing function represents a main function for a network in general and for Adhoc networks (i.e. wireless networks without infrastructure) more particularly. Routing protocols in these networks have been a subject of numerous researches; several approaches have been discussed, and many protocols [3] [4] have been proposed. The most cited (quoted) in literature is probably the AODV protocol [5]. Ensuring routing QoS consists in determining one or several (paths) that satisfy best QoS constraints such as packet loss, throughput, jitter, etc.
In this paper a new variant M-AODV (M for Modified) that discovers in a first step, all possible paths between sources and destinations and maintain them during all data transfer phase. In case of a failure of the actual route, the data transfer will use one of the previously established routes (secondary routes). The failure state is declared only if all paths, found in discovery phase, cannot be used.
In this study, we focus on QoS metrics such as load control (overhead), reliability (packet loss), the packets delay transit etc... under various constraints like mobility, energy and scaling from witch suffers the majority of routing algorithms in MANETs.
The remainder of this paper is organized as follows: in section 2, we give a brief review of QoS. In section 3, we discuss the most important characteristics of the AODV protocol, in section 4, we present our new protocol variant and the proposed changes. We evaluate later in section 5, the performance of this new AODV variant by simulation using NS-2 (Network Simulator), considering several contexts. We finish with conclusion and future recommendations of our researches.

## 2. Quality of Service (QoS) in the MANET

2.1. Qos model

Quality of Service (QoS) refers to a set of mechanisms able to share fairly various resources offered by the network to each application as needed, to provide, if possible, to every application the desired quality (the network's ability to provide a service) [6].

The QoS is characterized by a certain number of parameters (throughput, latency, jitter and loss, etc.) and it can be defined as the degree of user satisfaction.

QoS model defines architecture that will provide the possible best service. This model must take into consideration all challenges imposed by Ad-hoc networks, like network topology change due to the mobility of its nodes, constraints of reliability and energy consumption, so it describes a set of services that allow users to select a number of safeguards (guarantees) that govern such properties as time, reliability, etc.. [7][8].

Classical models like Intserv / RSVP [9] and DiffServ [10] proposed in first wired network types are not suitable (adapted) for MANETs. Various solutions or models [11] [12] namely: 2LqoS (Two-Layered Quality), CEDAR, noise, FQMM (Flexible QoS Model for MANET), SWAN (Service Differentiation in Wireless Ad-hoc Networks) and INSIGNIA have been proposed for the Ad-hoc networks.

Each of these models attempts (tries) to improve one or several QoS parameters, as they may be part of one or more network layers architecture.

2.2. QoS Routing

New requirements (needs) for multimedia and real-time applications require few delay and very high data rates which require (oblige) the use of new routing protocols supporting QoS [13] [14].

The QoS support must take in consideration a number of Ad-hoc networks constraints (mobility, energy, scale, etc.). QoS can be introduced into different layers network if there is need (channel access functions at MAC layer, routing protocols at network layer, etc.).[15].

Routing operation consists to find routes between communicating entities (transmitter / receiver) able to convey data packets continuously using less bandwidth and fewer packets control. Routing in MANETs must also manage constraints of nodes energy problems, topology frequent changes due to nodes mobility and communication channel nature (air). QoS routing can be defined as the research for routes satisfying the wanted (desired) QoS. To be as eligible routes, they must satisfy a number of constraints (such that delay, bandwidth, reliability, etc.) [16]. Indeed, any path that satisfies a number of quantitative or qualitative criteria can be described as path providing (ensuring) certain QoS.

## 3. Original AODV

3.1 Introduction

The AODV protocol (Ad-hoc on demand Distance Vector) [7] is a reactive routing protocol based on the distance vector Principle, combining unicast and multicast routing.

In AODV, the path between two nodes is calculated when needed (if necessary), i.e. when a source node wants to send data packets to a destination, it finds a path (Discovery Phase), uses it during the transfer phase, and it must maintain this path during its utilisation (Maintenance Phase).

The finding and maintaining process of a path is based on the exchange of a set of control packets: RREQ (Route REQueset), RREP (Route Reply), RERR (Route Error), RRepAck (Route Reply Acknowledgment) and Hello messages (Hello). RREQ is initiated by the source node to find a path in multicast mode. RREP is used by an intermediate or destination node to respond to a request of path finding in unicast mode. Hello messages are used to maintain the consistency of a previously established path.

Routing table is associated for each node in AODV protocol with containing: the destination address, the list of active neighbors, the number of hops (hop) to reach the destination, time of expiration after which the entry is invalidated, and so on.

To avoid the formation of infinite loop, AODV uses the principle of sequence numbers, limiting the unnecessary transmission of control packets (problem of the overhead); these numbers allow the use of fresh routes following the mobility of nodes, as they ensure the coherence and consistency of routing information [5].

It should be noted when the path breaks due to the absence of one node either by removal or a problem of energy, a local repair procedure (local repair) is called, it takes over the reconstruction of the path from this point. If this procedure cannot solve the problem, the source node try to find a new path and the number of attempts (RREQ_RETRIES) is decremented by 1, until the success or failure of the communication link.

This procedure generates a considerable amount of control packets. It should be noted that the original AODV maintains only one path to destination. To address this problem, it is preferable to have an alternative path already prepared. Two solutions are possible [8]: AODV with relief paths or multi-paths and there are two variants:
Several paths from source noted (M-AODV)
Several paths by intermediate node noted (M-AODV-I)

Paths from source or intermediate node are either completely disjoint (totally separated) or with common links.

**Completely disjoined paths**: a break at a route does not affect the rest of routes and therefore the use of another route is always possible to transmit data.

**Paths with common links**: sometimes, connection between two unspecified nodes belongs to several routes and when it break, all routes passing through this section are not in use and we have very small number of routes available adding to this we are obliged to generate very important additional traffic to notify source of this disconnection.

## 4. Proposed variant

### 4.1. Motivations

Among the key points having motivated our proposition (Modification of AODV protocol) we can cite: improving mechanisms that generate data packets loss (broken link or queues overflow associated at each nodes), the rational use of bandwidth (flow) and reducing packets latency.

Two cases are causing packet loss. The first one is due to frequent topology change by migration or remoteness (mobility) node formerly is part of link and its downstream and upstream neighbors respectively continues to send acknowledgments and data packets for a certain period before realize that link is failing (broken). The second is when a node starts the local repair procedure after detecting a broken link, the source was not aware of this situation, therefore continues to send its data packets normally, causing an overflow queue associated with the node without these data will be transmitted to their destination.

Loss can therefore be improved by changing the discovery and maintenance mechanisms of routes providing an almost continuous availability of links between communicating pairs (multi paths).

Rationalize the bandwidth use back to allow more useful data transfer and less control packets such as path discover (RREQ, RREP, etc.) path maintenance (Hello) which significantly reduces overload problem afflicting almost all wireless networks.

### 4.2. M-AODV protocol "M stands for Modified"

In this solution, we adds information to control packets for all routes and after exploring all possible paths, one with the shortest path hop count is first selected that respect QoS criterion required by user. In this solution control packet RREQ and RREP are routed in broadcast way. When the source wishes to transmit, it checks its routing table for any valid route to desired destination. If this is not the case, it starts Discovery Phase (discovery route process) by broadcasting control packet RREQ (Fig. 1).

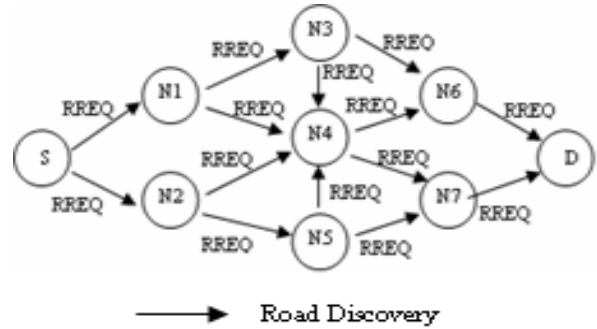

Fig. 1: Path discovery in M-AODV

As soon as an unspecified node has any path to destination, it responds to source with RREP packet and if isn't the case, destination respond to source by broadcasting a control packet RREP (Fig. 2) to trace (recall) all possible routes.

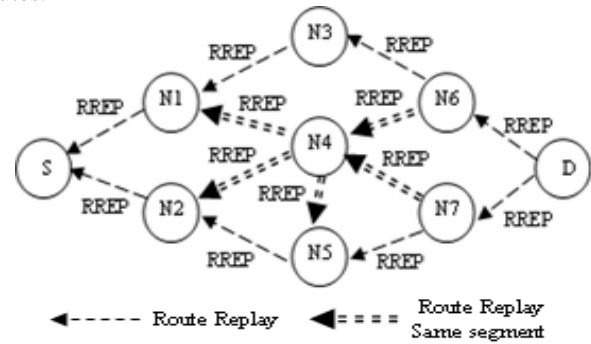

Fig. 2: Reverse Path

Different possible paths are: (S, N1, N3, N6, D) (S, N1, N4, N6, D) (S, N1, N4, N5, N7, D) (S, N1, N4 , N7, D) (S, N2, N5, N7, D) (S, N2, N4, N6, D) (S, N2, N5, N4, N6, D) and (S, N2, N5 , N4, N7, D). Once the node N4 is carried out, the different routes which pass (there) will not be valid.

Different Completely disjoined routes are selected from paths passed by low degree nodes: (S, N1, N3, N6, D) and (S, N2, N5, N4, N7, D), one will be taken as Primary path and other as secondary (minor) routes (Fig. 3).

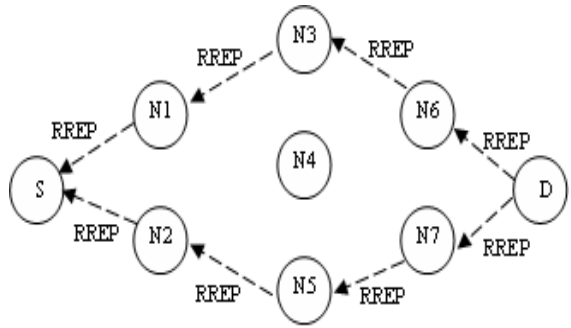

Fig. 3: Disjoint paths in M-AODV

Management process of routes is source responsibility but intermediate nodes are responsible only on their routing tables.

In link failure case, source stops transmission and repeats (reiterates) operation after selecting a new route from the spare paths available to it (minor routes).

In adopted (restraint) solution, the number of completely disjoined paths is fixed first to a number less than or equal to "n0" with a threshold of "s0". Once the threshold is reached in parallel with current data transfer, discovery phase for new route is initiated to determine other routes until reaching (to achieve) "n0". Although theoretically this solution generates a sizeable overhead but has the advantage of route availability at any time.

In original AODV when path is broken, local repair phase is initiated and if failure is declared a new discovery phase is initiated by source node.

In this modification, we propose to eliminate local repair phase to minimize modified protocol (M-AODV) task and the discovery phase is delegated in all scenarios to source node for a number of attempts RREQ_RETRIES (without M-AODV local repair).

Maintenance of all routes follows the same principle as original AODV by using "Hellos" packets (see Fig. 4).

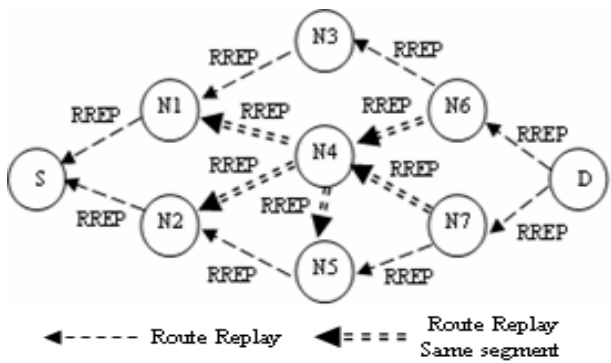

Fig. 4: Data transfer and maintenance phase

The path (S, N2, N5, N7, D) is considered as main street and (S, N1, N3, N6) as minor route or rescue.

## 5. Simulation

Measures taken in this simulation studies must be made on a set of parameters (metrics) like packet loss, load control, etc. and this must be observed under certain conditions (constraints) such as mobility, density, scale etc...

### 5.1. Constraints
#### 5.1.1. Mobility

Once route is built (established), it will be used and maintained for fixed period or until path is broken. In an Ad-hoc network, nodes are mobile, they will not be in neighbor's scope and therefore the route or routes which they are member become invalid. In this case, we try to revive (start again) discovery phase which generates an additional volume control packets [17].

The objective here is to test original AODV routing protocol behavior and modified variant M-AODV under constraint of mobility to know if amended (proposed) version minimizes control packets volume generated when establishing routes, transit delay of data packets and at the same time data packets loss or control packets transmitted.

#### 5.1.2. Energy

In general energy consumption is proportional to the number of packets processed and the type of treatment performed (carried) (transmission / reception), it is noted that packet emission requires more energy than reception [18]. Here the objective is to determine (know) which of the two protocols (AODV and M-AODV) manages in best manner nodes energy.

#### 5.1.3 Density

It means the number of nodes used or involved in an Adhoc networks. This constraint is examined to determine the impact of the nodes number on the network overload, how it is easy to determine routes if nodes number is so important, etc.

#### 5.1.4 Scale Transition (network topology)

It means space or scale and not the sense in usual networks topology (star, bus, ring ... etc…). Currently most routing protocols in MANETs suffer from the problem of scalability (from few meters to tens meters). The objective here is whether this constraint is well respected and which of the two protocols tested keep its performances face to scale transition.

### 5.2. Metrics
#### 5.2.1. Packet loss

Indeed, packet loss is crucial factor [16] to evaluate routing protocol performances. Ensure zero loss transfer is desired (coveted), but is that possible in Ad-hoc networks? A protocol is efficient (powerful), if it can minimize to maximum the packet loss in all conditions to which it is confronted (large or small nodes number (density), a small or large scale, high or low mobility ... etc.).

#### 5.2.2. The load control

As for packet loss parameter, load control is a crucial element [11] for protocol performance evaluation. Over control packet volume is large, performance degrades and more bandwidth is used by control packets than data. The measure of such parameter can justify the choice of using such or such protocol.

#### 5.2.3. Rate (bandwidth)

Is the maximum information quantity per unit time? It's actually the maximum transfer ratio can be maintained between transmitter and receiver, this factor depends on physical links and also on flows sharing them. A better bandwidth [16] [19] management allows to pass high rate is for such reasons it is essential to measure our proposed protocol performance on this parameter.

To study and analyze our proposal behavior, we used the Network Simulator (Ns2) [20] version 2.31 installed on Debian GNU / Linux. Simulation context consists of 20 nodes in a region 800x600m2. Transmission range is 250m for a perfect space (unobstructed) with Random Way Point (RWP) mobility model. Nodes moves at a maximum speed of 5m / s (average speed). Traffic was automatically generated randomly using Ns2 script cbrgen.tcl for Constant Bit Rate (CBR) of 512 bytes according to UDP protocol. Different mobility scenarios are also produced with Ns2 Setdest program. Time Simulation is set to 120 s for all test and each node has 10 joules as initial value.

### 5.3. Curves & discussions

The desired parameters to be evaluated by simulation under different contexts (mobility, density, etc.) are: throughput in kbps that indicates data transfer rate. Having a network system with high flow is coveted. Average end-to-end delay (e2e) reflects time taken between data packet transmission and reception. More time is short over the network is requested. Packet Delivery Ratio (PDR) describes the ratio between successfully delivered packets and the total number of transmitted packets. More the value of the PDR is small more the network is effective. Normalized Routing Load or Normalized Overhead Load (NRL or NOL) is just the ratio of transmitted packets control on received packets number. This value expresses overhead network. Packets lost ratio is the rapport among successfully received and sent packets. This ration proves network reliability. The last parameter is the consumed energy by each node in whole network or only in routing phase to see which algorithms manage better this resource.

In the following stage, we will study the mobility impact (respectively: nodes number) on the above parameters such as a high mobility (pause time equal to zero) to no mobility state (for pause time equal to 200s and more). (Respectively: by varying network size from small network with 10 nodes to denser network of 100 nodes).

Figures (Fig. 5 & 6) show that the presence of backup paths (multiple paths) in M-AODV version improves a better throughput especially for a high and medium mobility and network size less than 80 nodes; while delay in basic version (AODV) takes over for high mobility (pause time less or equal to 80 seconds) but beyond this value M-AODV takes the hand (Fig. 7). The same remark is done for network size ranging from 40 to 80 nodes (Fig. 8). Packet loss ratio is smaller in M-AODV for low and medium mobility due to available routes number and network size up to 75 nodes (Fig.9 & Fig.10). The PDR for our proposal method shows acceptable values for medium network size above 70 nodes (Fig. 12) and for low and medium mobility (Fig. 10). NRL ratio is more or less balanced i.e. sometimes is high due to generation of more packets control in both variants (more paths in M-AODV and more discovery operations in AODV) (Fig.13 & Fig.14). For consumed energy over time by all nodes is almost identical in either AODV or M-AODV (Fig. 15) but M-AODV minimizes this resource better than in routing phase (Fig .16)

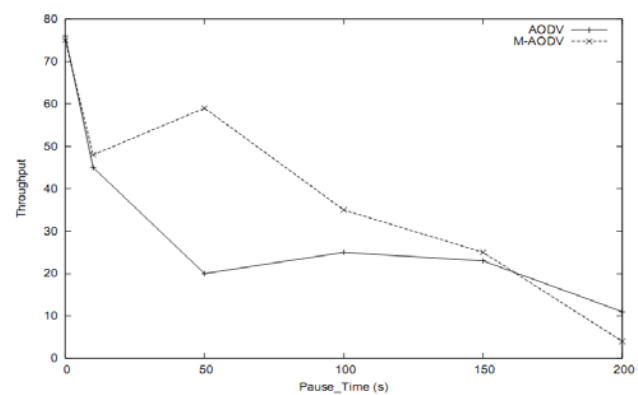

Fig. 5 Throughput Vs Pause_Time

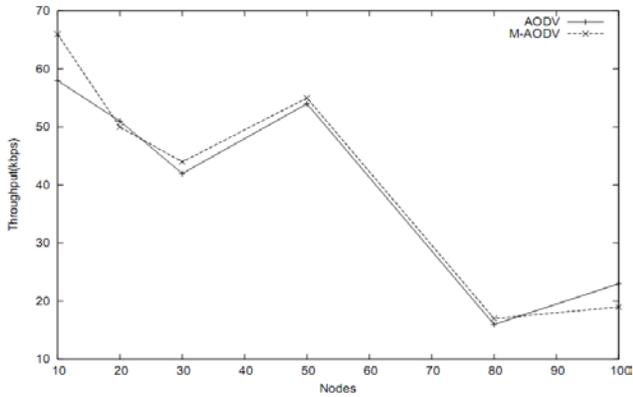

Fig. 6 Throughput Vs Nodes

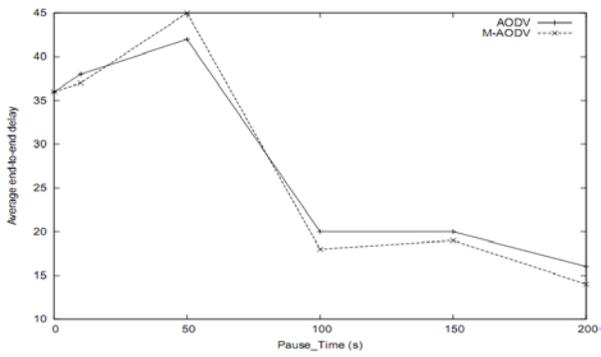

Fig. 7 Average_end_to_end_delay Vs Pause_Time

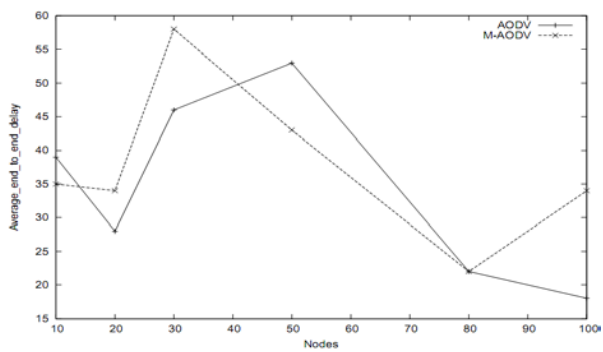

Fig. 8 Average_end_to_end_delay Vs Nodes

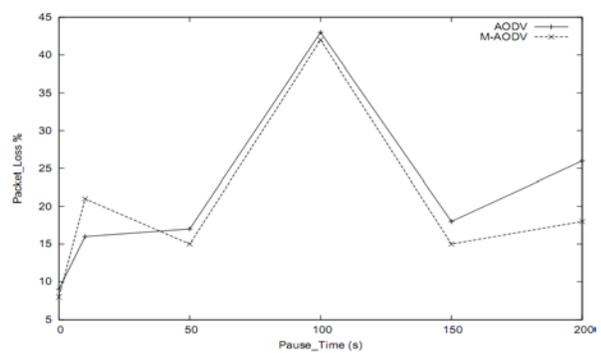

Fig. 9 Packet_Loss Ratio Vs Pause_Time

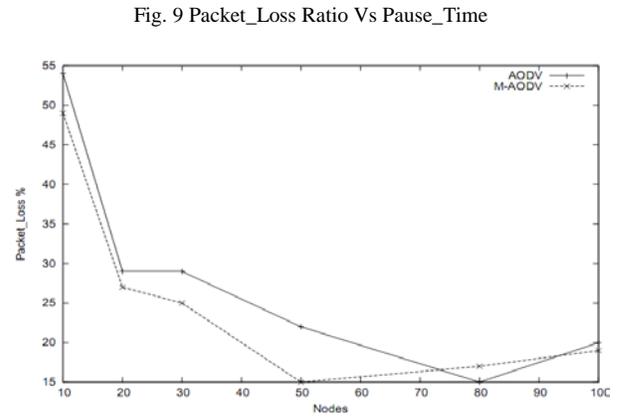

Fig. 10 Packet_Loss Ratio Vs Nodes

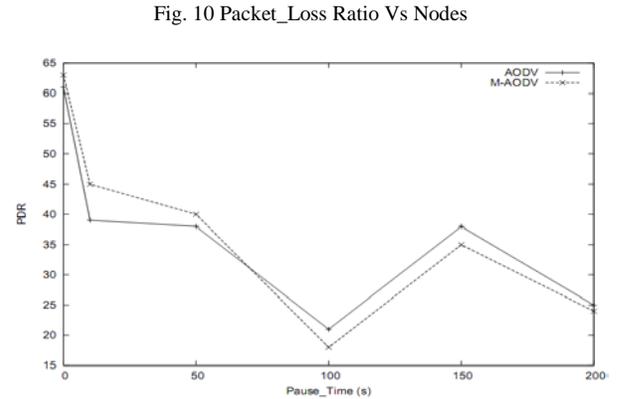

Fig. 11 Packet Delivery Ratio Vs Pause_Time

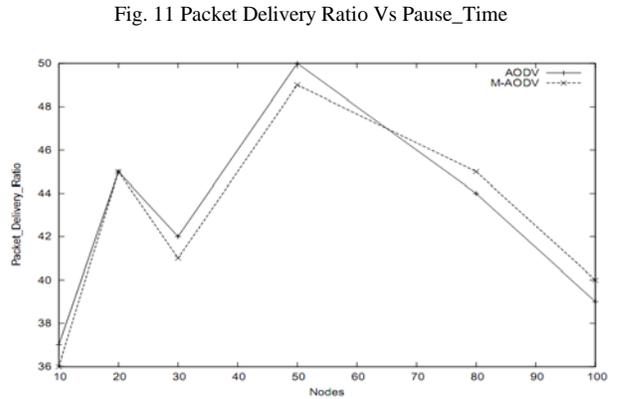

Fig. 12 Packet Delivery Ratio Vs Nodes

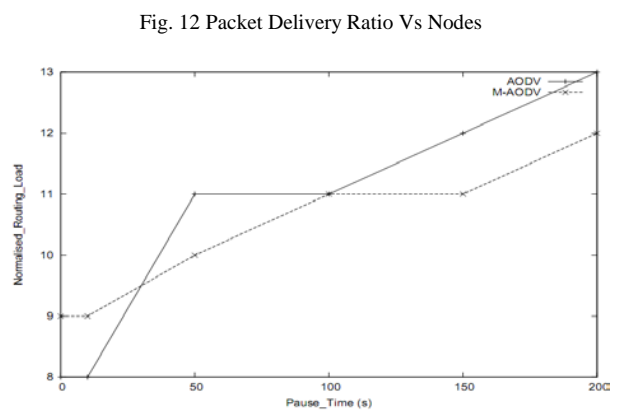

Fig. 13 Normalised_Routing_Load Vs Pause_Time

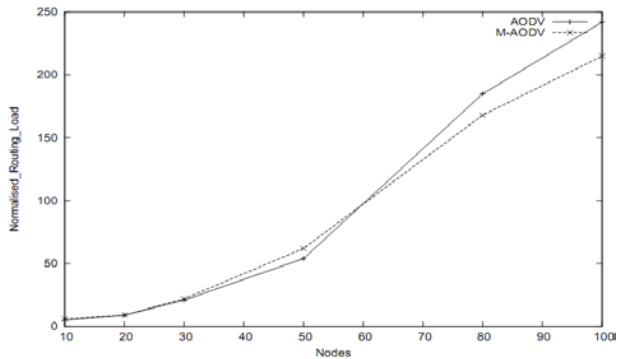

Fig. 14 Normalised_Routing_Load Vs Nodes

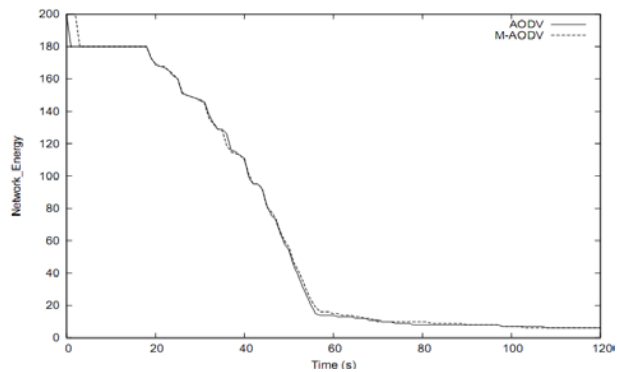

Fig. 15 Network_Energy Vs Time

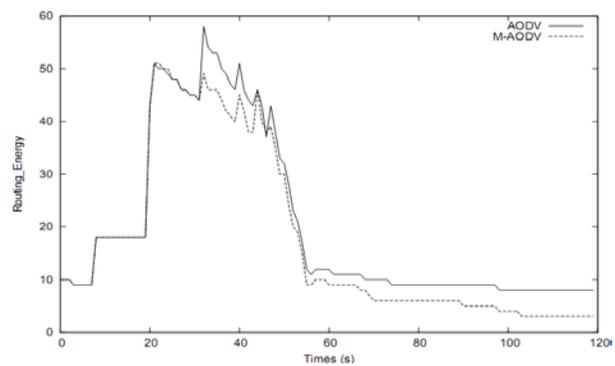

Fig. 16 Routing_Energy Vs Time

## 6. Conclusion & Perspectives

The proposed M-AODV generates in a first time, a number of possible paths (multiple paths) between sources and destinations with the aim to use them if needed (to minimize the repair phase overhead) i.e. when the link in use is broken; by running discovery phase in parallel with data packets transfer phase ensure to have a number of adjacent paths in advance. The performed simulation and comparison between the original AODV protocol and M-AODV show that this last can improve the QoS in Mobile ad hoc networks under different conditions.

Future extensions of AODV protocol should be based on the prediction of a future link disconnection considering the signal quality and mobile node speed. We project also to extend the functionality of the protocol with the aim of adapting it to large scale networks.